\documentclass[preprint,aps]{revtex4}
\usepackage{amsmath,amsfonts,amssymb,bm,txfonts}
\usepackage{graphicx}
\usepackage{color}

\newcommand{\cL}{\mathcal{L}}

\newcommand{\cD}{\mathcal{D}}
\newcommand{\cP}{\mathcal{P}}
\newcommand{\cQ}{\mathcal{Q}}
\newcommand{\Ker}{\mathtt{Ker\,}}
\newcommand{\tIm}{\mathtt{Im\,}}
\newcommand{\Span}{\mathtt{Span\,}}

\begin{document}
\title{Origin of Temperature Gradient in Nonequilibrium Steady States
in Weakly Coupled Quantum Spin Systems}

\author{Toyohiko Ishida and Ayumu Sugita}
\email{sugita@a-phys.eng.osaka-cu.ac.jp}
\affiliation{
Department of Applied Physics, Osaka City University,
3-3-138 Sugimoto, Sumiyoshi-ku, Osaka 558-8585, Japan
}
\date{\today}

\begin{abstract} 
We study nonequilibrium steady states (NESSs) in quantum 
spin-1/2 chains 
in contact with two heat baths at different temperatures. 
We consider the weak-coupling limit both for spin-spin coupling in the system
and for system-bath coupling. This setting allows us to treat NESSs 
with a nonzero temperature 
gradient analytically. 
We develop a perturbation theory for this weak-coupling situation
and show a simple condition 
for the existence of nonzero temperature gradient.
This condition is independent of the integrability of the system.
\end{abstract}

\maketitle

\section{Introduction}
The microscopic description and understanding of nonequilibrium steady states
(NESSs) is one of the most challenging problems in physics.
For heat-conducting systems, Fourier's law states that the heat current is proportional to the temperature gradient. 
Although it is a very universal empirical law, there has been no microscopic 
derivation of it.

In a normal heat-conducting NESS, we observe a 
uniform nonzero temperature gradient and heat current. 
However, when considering a simple theoretical model,
we often observe a flat temperature profile in the system, which 
represents ballistic transport.
It is commonly believed that normal heat-conducting states
are realized in non-integrable systems.\cite{lepri} However, 
there are some exceptions to this rule.
\cite{prosen,michel,wu,mendoza} 
Therefore, the actual condition for realizing normal NESSs
is still an open problem.

\begin{figure}[ptb]
\begin{center}
\includegraphics[width=14cm]{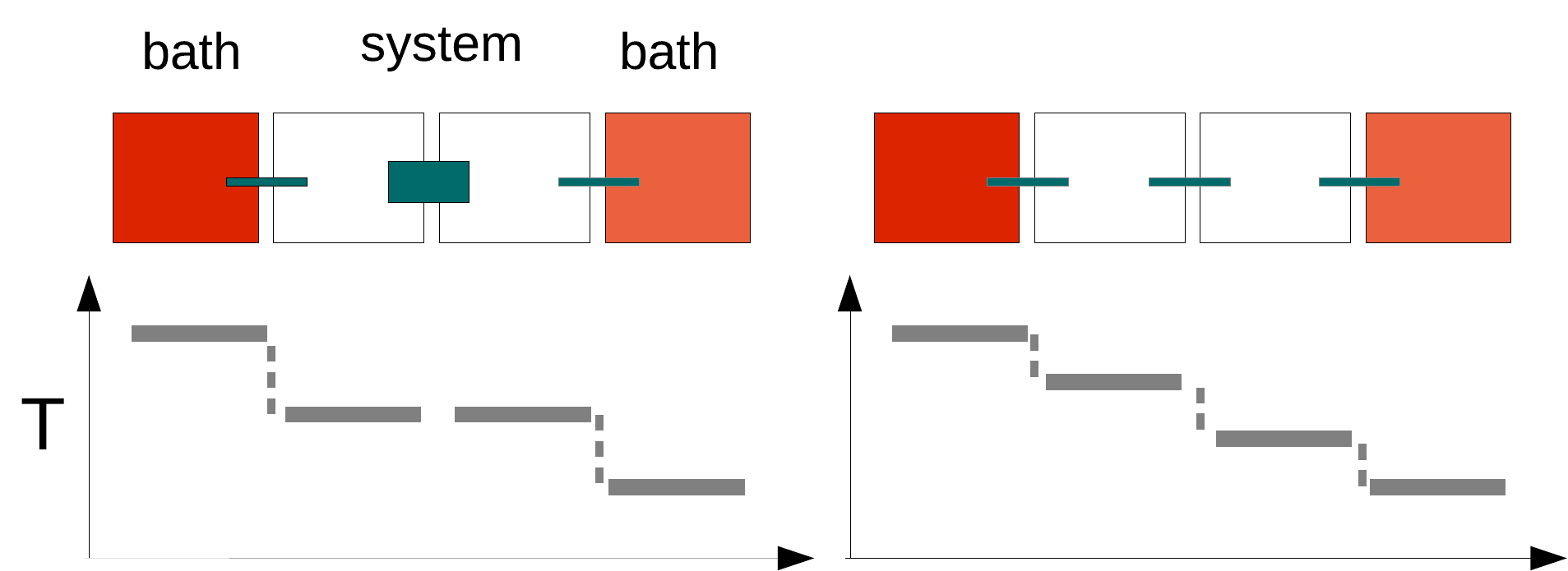}
\caption{When the system-bath coupling is weak but the coupling 
inside the system is not weak, we expect a flat temperature profile
because the thermal resistance is large only at the boundaries. If
the coupling in the system is also weak, we expect a nonzero temperature
gradient in the system.}
\label{weakcoupling}
\end{center}
\end{figure}

In this paper, we consider heat-conducting NESSs in 
one-dimensional quantum spin chains in contact with
two heat baths. In the study of  such NESSs, the coupling between 
the system and the baths is often assumed to be weak.
\cite{yuge,sugita}
However, if the system-bath coupling is weak and the spin-spin
interaction is not weak, the system will have a flat temperature
profile because the thermal resistance at the boundary is large
compared with that in the system (Fig. \ref{weakcoupling}). 
To study NESSs with nonzero
temperature gradients, we consider the weak-coupling limit both 
for spin-bath coupling and for spin-spin coupling. 
In this limit, the thermal resistance uniformly becomes infinitely large 
both in
the system and at the boundary. Therefore, we
expect a ``frozen'' NESS to appear, 
which has a nonzero temperature gradient
and no current. We will show that the ``frozen'' solution 
is not just a tensor product of local equilibrium states. 

Heat-conducting states with nonzero temperature gradients
are difficult to study 
analytically, because they are usually realized in
non-integrable systems. However, our setting allows 
some analytical treatment of such states. In this paper,
we consider one-dimensional uniform quantum spin-1/2 
systems with nearest-neighbor interactions, whose Hamiltonians
are of the form
\begin{eqnarray}
H &=& \lambda \sum_l \sum_{j,k}\alpha_{j,k}\sigma_l^j\sigma_{l+1}^k
 + h\sum_l\sigma_l^z.
\end{eqnarray}
Here, $\sigma_l^j\; (j=x,y,z)$ represents a Pauli matrix at the $l$th site.
We assume $h\ne 0$ and $\alpha_{ij} = \alpha_{ji}$.
Then we show that 
\begin{eqnarray}
\alpha_{zz} \ne 0
\label{alphazz}
\end{eqnarray}
is necessary to form a nonzero temperature gradient.
Note that this condition is independent of the integrability 
of the system. For example, it is consistent with the observation
that the $XXZ$ model has normal conducting states in spite of
its integrability\cite{prosen, michel, wu, mendoza}. 
We can understand intuitively
that Eq. (\ref{alphazz}) must be satisfied
so that the local temperature at a site can affect
the temperatures of neighboring sites,
because in the weak-coupling limit the local temperature
is determined only from $\langle\sigma_z\rangle$.
We also show that 
some three-body correlations play important roles in forming the
temperature gradient. The forms of the correlation terms are universal,
i.e., they do not depend on the form of the interaction.

This paper is organized as follows. In Sect. \ref{QMEsection}, we derive
the quantum master equation (QME) in the weak-coupling limit. We show that
the heat bath superoperator becomes local when the
spin-spin coupling is weak. In Sect. \ref{perturbation}, we derive
a perturbative expansion of the stationary solution of the
QME. In Sect. \ref{spinchain} we study quantum spin-1/2
chains. First we analyze two-spin cases and show an explicit
example of a frozen NESS with a nonzero temperature gradient. Then
we study $N$-spin systems and
derive Eq. (\ref{alphazz}). We also show some numerical
results to verify our theoretical considerations.
Section \ref{summary} is 
devoted to a summary.

\section{Quantum Master Equation}
\label{QMEsection}
In this section, we derive the QME for
weakly coupled spin systems.
For simplicity, we first derive the QME for
a system coupled to a single heat bath. Generalization to 
multiple heat baths is straightforward.
We denote the Hilbert space of the system by $\mathsf{H}$
and assume that its dimension is finite.
The set of all linear operators on $\mathsf{H}$ is denoted by 
$\mathsf{L}$. It becomes a Hilbert space
by introducing the Hilbert--Schmidt inner product
as ${\rm Tr}\left(X_1 X_2\right)$ for any $X_1, X_2 \in \mathsf{L}$.
We refer to a linear operator on $\mathsf{L}$ 
as a superoperator.

We start with the Hamiltonian
\begin{eqnarray}
H &=& H_{\rm S} + H_{\rm B} + \lambda H_{\rm SB},
\end{eqnarray}
where $\lambda$ is a small coupling parameter, 
$H_{\rm S}$ is the Hamiltonian for the system, and $H_{\rm B}$ is that for the
heat bath. $H_{\rm SB}$ represents the interaction between the
system and the heat bath, which can be 
written as
\begin{eqnarray}
H_{\rm SB} &=& \sum_j X_j Y_j.
\end{eqnarray} 
Here, $X_j$ and $Y_j$ are Hermitian operators on 
the system and on the heat bah, respectively.
By using the Born-Markov approximation\cite{kubo, breuer}, we obtain
the standard Redfield-type QME
\begin{eqnarray}
\frac{d}{dt}\rho(t) &=& 
\frac{1}{i\hbar}\left[H_{\rm S}, \rho(t)\right] 
+ \lambda\frac{1}{i\hbar}\left[\overline{H}_{\rm SB}, \rho(t)\right] 
+ \lambda^2\cD\rho.
\label{QME}
\end{eqnarray}
Here, $\overline{H}_{\rm SB}\in \mathsf{L}$ represents the 
interaction averaged over the heat bath, whose
explicit form is 
\begin{eqnarray}
\overline{H}_{SB} &=& {\rm Tr}_{\rm B} H_{\rm SB} = \sum_j X_j \overline{Y}_j,
\end{eqnarray}
where $\overline{Y}_j = {\rm Tr}\left\{\rho_{\rm B} Y_j\right\}$ and
$\rho_{\rm B}$ is the equilibrium state of the heat bath $\rho_{\rm B} = e^{-\beta H_{\rm B}}/Z_{\rm B}$.
$\cD$ is the heat bath superoperator
\begin{eqnarray}
\cD \rho &=& - \frac{1}{\hbar^2}\sum_{jl}\left[X_j, 
S_{jl}\rho\right] + {\rm h.c.},
\label{D}
\end{eqnarray}
where
\begin{eqnarray}
S_{jl} &=& 
\int_0^\infty dt e^{-iH_{\rm S}t/\hbar}X_l e^{iH_{\rm S}t/\hbar}\Phi_{jl}(t).
\end{eqnarray}
$\Phi_{jl}$ is a correlation function for the heat bath,
\begin{eqnarray}
\Phi_{jl}(t) &=& {\rm Tr}
\left\{\rho_{\rm B} \Delta Y_j(t) \Delta Y_l \right\},
\end{eqnarray}
where $\Delta Y_k(t) = Y_k(t) - \overline{Y}_k $.
Note that Eq. (\ref{QME}) is correct up to $O(\lambda^2)$.
We also use the Markovian approximation in deriving Eq. (\ref{QME}),
which is correct when the density matrix in the interacting picture
$e^{iH_{\rm S}t/\hbar}\rho(t)e^{-iH_{\rm S}t/\hbar}$ is slowly varying
compared with the correlation time of the heat bath \cite{kubo}.
This assumption holds for NESSs in a weakly nonequilibrium regime.
In our perturbation theory, the zeroth order solution satisfies
this condition exactly even when it is far from equilibrium.

\begin{figure}[ptb]
\begin{center}
\includegraphics[width=12cm]{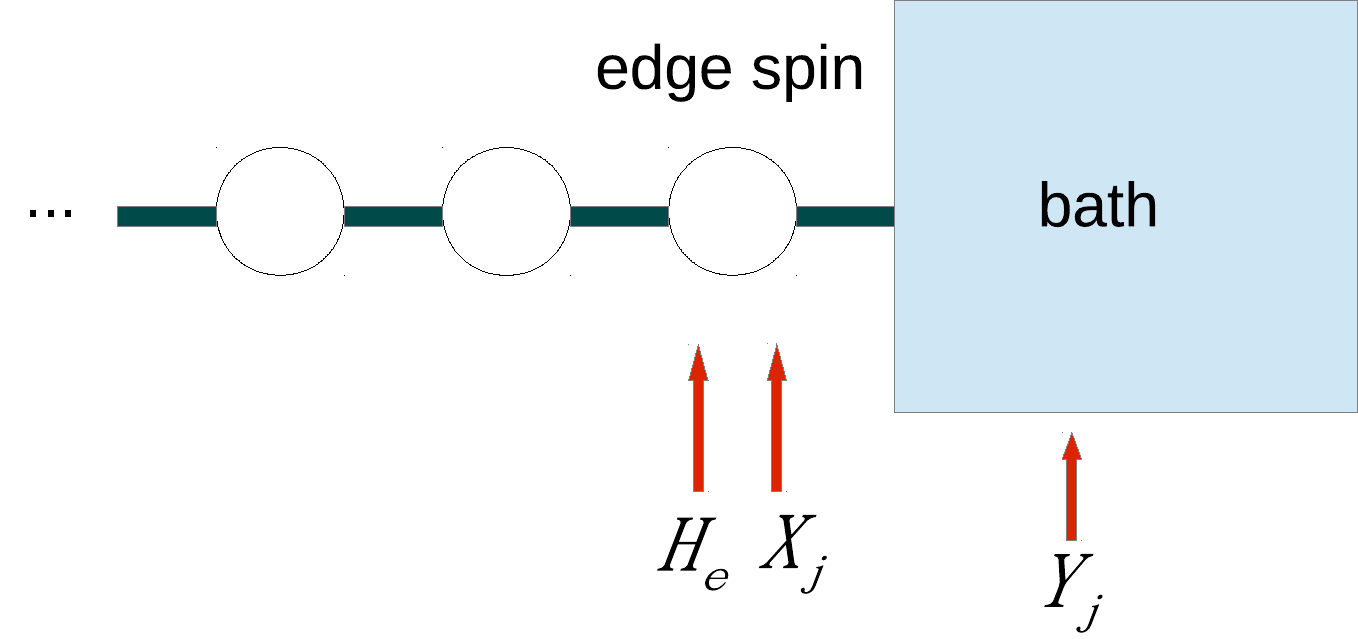}
\caption{$X_j$ and $H_{\rm e}$ act only on the edge spin.}
\label{edge_spin}
\end{center}
\end{figure}
Now we consider a weakly coupled one-dimensional spin system 
with nearest-neighbor interactions. We assume that the spin-spin
interaction is of the same order as the system-bath interaction.
Then the system Hamiltonian 
can be written as
\begin{eqnarray}
H_{\rm S} &=& H_0 + \lambda H_{\rm I}, 
\end{eqnarray}
where $H_0$ represents the external field applied to the spins
and $H_{\rm I}$ is the interaction among the spins. We also assume that
the $X_j$ in $H_{\rm SB}$ act only on the edge spin (Fig. \ref{edge_spin}) 
and write
\begin{eqnarray}
H_0 &= & H_{0}' + H_{\rm e},
\end{eqnarray} 
where $H_{\rm e}$ acts on the edge spin and $H_{0}'$ acts on the other spins.
Since $[H_{0}', X_j] = 0$, we have
\begin{eqnarray}
S_{jl} &=& \int_0^\infty dt e^{-iH_{\rm e} t/\hbar}X_l e^{iH_{\rm e} t/\hbar}\Phi_{jl}(t) + O(\lambda).
\end{eqnarray}
The $O(\lambda)$ term is negligible in the QME in Eq. (\ref{QME}),
and now $S_{jl}$ is a local operator acting on the edge spin. 
It can also be written as
\begin{eqnarray}
S_{jl} &=& \sum_{pq}\Xi_{jl}\left(\frac{E_p-E_q}{\hbar}\right)
|E_p\rangle \langle E_p|X_l|E_q\rangle \langle E_q|,
\label{S}
\end{eqnarray} 
where $\Xi_{jl}$ is the Fourier--Laplace transform of the correlation function
\begin{eqnarray}
\Xi_{jl}(\omega) &=&
\int_0^\infty e^{-i\omega t}\Phi_{jl}(t),
\end{eqnarray}
$E_r$ is an eigenvalue of $H_{\rm e}$ and $|E_r\rangle$ is the corresponding
eigenvector.

If we place heat baths at both ends of the spin chain,
the QME for this weak-coupling limit is
\begin{eqnarray}
\frac{d}{dt} \rho &=& \cL\rho, 
\end{eqnarray}
\begin{eqnarray}
 \cL &=& \cL_0 + \lambda \cL_1 + \lambda^2 \cL_2,
  \label{liouvillian}
\end{eqnarray}
\begin{eqnarray}
\cL_0\rho &=& \frac{1}{i\hbar}[H_0, \rho],\\
\cL_1 \rho &=& \frac{1}{i\hbar}[H_1, \rho],\\
\cL_2 \rho &=& \cD_{\rm L} \rho + \cD_{\rm R} \rho,
\end{eqnarray}
where $H_1 = H_{\rm I} + \overline{H}_{\rm SB}$, and
$\cD_{\rm L}$ and $\cD_{\rm R}$ are
heat bath superoperators for the left and right heat baths, respectively.
They are of the form given in Eqs. (\ref{D}) and (\ref{S}).

\section{Perturbative Expansion}
\label{perturbation}

\subsection{Zeroth-order solution}

The steady state is determined by the equation
\begin{eqnarray}
\frac{d}{dt}\rho = \cL\rho = 0. 
\label{ness_eq}
\end{eqnarray}
We calculate the steady state by expanding $\rho$ with respect to
the coupling parameter $\lambda$:
\begin{eqnarray}
\rho &=& \rho_0 + \lambda \rho_1 + \lambda^2 \rho_2 + \dots .
\end{eqnarray}
By substituting it into Eq. (\ref{ness_eq}), we obtain
the following equations up to the second order:
\begin{eqnarray}
\cL_0 \rho_0 &=& 0,\label{0th}\\
\cL_0 \rho_1 + \cL_1 \rho_0 &=& 0,\label{1st}\\
\cL_0 \rho_2 + \cL_1 \rho_1 + \cL_2 \rho_0 &=& 0.\label{2nd}
\end{eqnarray}

In usual perturbation theories, the zeroth-order solution is determined
from the zeroth-order equation. In this case, however, 
we cannot determine $\rho_0$ simply from Eq. (\ref{0th}),
and the first order-equation Eq. 
(\ref{1st}) also fails to determine $\rho_0$. Hence we
need all three equations Eqs. (\ref{0th})-(\ref{2nd}) to determine
the zeroth-order solution. Physically speaking, this is 
because the Liouvillian $\cL_0 + \lambda \cL_1$ represents
Hamiltonian dynamics governed by $H_0 + \lambda H_1$,
 and all energy eigenstates of 
$H_0 + \lambda H_1$ are stationary in this dynamics. 
We require the heat baths, which are represented by $\cL_2$, to make the
stationary state unique.

Let us consider the eigenvalue problem of the Hamiltonian
$H_0 + \lambda H_1$ up to the first order of $\lambda$. 
Since $H_0$ is a non-interacting Hamiltonian
and highly degenerate, we must use degenerate perturbation
theory. Then we obtain the eigenvalues
\begin{eqnarray}
E_{n,k} &=& E_n + \lambda e^{(n)}_k, 
\end{eqnarray}
where $E_n$ is an eigenvalue of $H_0$ and $e^{(n)}_k$
is obtained by diagonalizing $H_1$ in the eigenspace
belonging to $E_n$. Therefore, the corresponding eigenvector 
$|E_{n,k}\rangle$ satisfies
\begin{eqnarray}
H_0 |E_{n,k}\rangle &=& E_n |E_{n,k}\rangle
\end{eqnarray} 
and
\begin{eqnarray}
\langle E_{n,k}|H_1 |E_{n,l}\rangle &=& e^{(n)}_k \delta_{k,l}.
\end{eqnarray}
We assume the $E_{n,k}$ are 
nondegenerate. 

Since $|E_{n,k}\rangle \langle E_{m,l}|\in \mathsf{L}$
satisfies the equation
\begin{eqnarray}
\cL_0 |E_{n,k}\rangle \langle E_{m,l}| 
&=& 
\frac{E_{n} - E_m}{i\hbar}
|E_{n,k}\rangle \langle E_{m,l}|, 
\end{eqnarray}
it is an eigenoperator of
$\cL_0$ with the eigenvalue $\frac{E_n - E_m}{i\hbar}$, and 
the operators of this type form a complete set
in $\mathsf{L}$.
The zero eigenspace is spanned by the operators with $n=m$:
\begin{eqnarray}
\Ker \cL_0 &=& \tt{Span}\left\{|E_{n,k}\rangle\langle E_{n,l}|\right\}_{n,k,l}.
\end{eqnarray}
We denote the projection superoperator to $\Ker\cL_0$ by $\cP_0$. 
Then we have
\begin{eqnarray}
\cP_0 \cL_0 = \cL_0 \cP_0 = 0
\end{eqnarray}
and 
\begin{eqnarray}
\cP_0 \rho_0 = \rho_0
\end{eqnarray}
because $\rho_0 \in \Ker\cL_0$.
By applying $\cP_0$ to Eq. (\ref{1st}), we obtain
\begin{eqnarray}
\cP_0 \cL_1 \rho_0 = \cP_0 \cL_1 \cP_0 \rho_0 = 0.
\end{eqnarray}
Hence, $\rho_0$ is in $\Ker\left(\cP_0 \cL_1 \cP_0\right)$,
which is spanned by the diagonal
elements:
\begin{eqnarray}
\Ker\left(\cP_0\cL_1\cP_0\right) &=& 
\tt{Span}\left\{ |E_{n,k}\rangle \langle E_{n,k}| \right\}_{n,k}.
\end{eqnarray}
We denote the projection superoperator to $\Ker\left(\cP_0\cL_1\cP_0\right)$
by $\cP_1$. Then
\begin{eqnarray}
\cP_1 \rho_0 = \rho_0.
\end{eqnarray}
We also define projection superoperators 
\begin{eqnarray}
\cQ_0 &\coloneqq& 1 - \cP_0, \\
\cQ_1 &\coloneqq& \cP_0- \cP_1.
\end{eqnarray}
Note that
\begin{eqnarray}
1 = \cQ_0 + \cQ_1 + \cP_1 
\end{eqnarray}
and
$
\tIm \cP_1 \subset \tIm \cP_0
$,
$
\tIm \cQ_1 \subset \tIm \cP_0
$.

Applying $\cQ_0$ to Eq. (\ref{1st}), we obtain
\begin{eqnarray}
Q_0 \cL_0 (P_0 + Q_0) \rho_1 + Q_0 \cL_1\rho_0
=
Q_0 \cL_0 Q_0 \rho_1 + Q_0 \cL_1\rho_0 = 0.
\end{eqnarray} 
Hence,
\begin{eqnarray}
\cQ_0\rho_1 &=& -\left(Q_0 \cL_0 Q_0 \right)^{-1} \cL_1 \rho_0.
\label{Q0rho1}
\end{eqnarray}
Note that the superoperator $\left(Q_0 \cL_0 Q_0 \right)^{-1}$ is
defined on ${\tt Im}Q_0$. We implicitly assumed in Eq. (\ref{Q0rho1}) 
that 
$\left(Q_0 \cL_0 Q_0 \right)^{-1}\rho=0$
if $\rho \notin {\tt Im}Q_0$. Hence, 
\begin{eqnarray}
\left(Q_0 \cL_0 Q_0 \right)^{-1} 
= 
\left(Q_0 \cL_0 Q_0 \right)^{-1}Q_0. 
\end{eqnarray}

By applying $\cP_1$ to Eq. (\ref{2nd}), 
we obtain
\begin{eqnarray}
\cP_1 \cL_1 \rho_1 + \cP_1 \cL_2\rho_0 = 0.
\end{eqnarray}
Since $\cP_1\cL_1\cP_0 = \cP_1\cP_0\cL_1\cP_0 = 0$, we have
\begin{eqnarray}
\cP_1 \cL_1 \rho_1 
&=& 
\cP_1 \cP_0 \cL_1(\cP_0 + \cQ_0) \rho_1\\
&=&
-\cP_1 \cL_1 \left(Q_0 \cL_0 Q_0 \right)^{-1} \cL_1 \cP_1 \rho_0.
\end{eqnarray}
We can show that
\begin{eqnarray}
\cP_1 \cL_1 \left(Q_0 \cL_0 Q_0 \right)^{-1} \cL_1 \cP_1 = 0
\label{mendou}
\end{eqnarray}
by a straightforward algebra (see Appendix\ref{proof_mendou}). 
Hence, we obtain the following three equations to determine the zeroth-order
solution:
\begin{eqnarray}
\cL_0 \rho_0 &=& 0, \\
\cP_0 \cL_1 \rho_0 &=& 0 \label{p0l1},\\
\cP_1 \cL_2 \rho_0 &=& 0.
\end{eqnarray}

\subsection{Higher-order solutions}
In a similar way, we can obtain higher-order solutions.
If the Liouvillian is expanded with respect to $\lambda$ as
\begin{eqnarray}
\cL &=& \sum_{n=0}^\infty \lambda^n \cL_n, 
\end{eqnarray}
the $m$th-order equation is
\begin{eqnarray}
\sum_{n=0}^m \cL_n \rho_{m-n} &=& 0. 
\label{mth_eq}
\end{eqnarray}
We decompose the $m$th-order solution as
\begin{eqnarray}
\rho_m &=& \cQ_0 \rho_m + \cQ_1 \rho_m + \cP_1 \rho_m
\end{eqnarray}
and represent the three terms by lower-order solutions
in the following. 

First we apply $\cQ_0$ to Eq. (\ref{mth_eq}). Then we have
\begin{eqnarray}
\cQ_0 \cL_0 \rho_m &=& - Q_0 \sum_{n=1}^m \cL_n \rho_{m-n}.
\end{eqnarray}
Hence, 
\begin{eqnarray}
\cQ_0 \rho_m &=& 
- \left(Q_0 \cL_0 Q_0\right)^{-1}
\sum_{n=1}^m \cL_n \rho_{m-n}.
\label{q0rhom}
\end{eqnarray}
Then we apply $\cQ_1$ to the $(m+1)$th-order 
equation. Since $\cQ_1\cL_0 = 0$, we
have
\begin{eqnarray}
\cQ_1 \cL_1 \rho_{m} &=& - \cQ_1 \sum_{n=2}^{m+1}\cL_n \rho_{m+1-n}.
\end{eqnarray} 
Because
\begin{eqnarray}
Q_1\cL_1  = Q_1\cL_1(P_1 + Q_1 + Q_0) = Q_1\cL_1(Q_1 + Q_0),  
\end{eqnarray}
we obtain
\begin{eqnarray}
Q_1 \rho_m &=& - \left(Q_1\cL_1 Q_1\right)^{-1} 
\left(\cL_1 Q_0 \rho_m + \sum_{n=2}^{m+1}\cL_n \rho_{m+1-n}
\right).
\label{q1rhom}
\end{eqnarray}

Next we apply $\cP_1$ to the $(m+2)$th-order equation. Then we have
\begin{eqnarray}
\cP_1 \cL_1 \rho_{m+1} + \cP_1 \cL_2 \rho_m 
+ \cP_1 \sum_{n=3}^{m+2} \cL_n \rho_{m+2-n} &=& 0.
\end{eqnarray}
The first term is
\begin{eqnarray}
\cP_1 \cL_1 (\cP_0 + \cQ_0)\rho_{m+1}
&=&
\cP_1 \cL_1 \cQ_0\rho_{m+1}\\
&=& 
- \cP_1 \cL_1 \left(Q_0 \cL_0 Q_0\right)^{-1}
\left(
\cL_1 \rho_m + \sum_{n=2}^{m+1}\cL_n\rho_{m+1-n}
\right)\\
&=&
- \cP_1 \cL_1 \left(Q_0 \cL_0 Q_0\right)^{-1}
\left(
\cL_1 (\cQ_1 + \cQ_0)\rho_m 
+ \sum_{n=2}^{m+1}\cL_n\rho_{m+1-n}
\right),
\end{eqnarray}
where we used 
Eq. (\ref{q0rhom}) with $m$ replaced by $m+1$ in the second line
and Eq. (\ref{mendou}) in the last line.
Therefore,
\begin{eqnarray}
\cP_1 \cL_2 \cP_1 \rho_m +
\cP_1 \left(\cL_2 - \cL_1 \left(Q_0 \cL_0 Q_0\right)^{-1}\cL_1\right)
(\cQ_1 + \cQ_0)\rho_m && \\
+ 
\cP_1 
\sum_{n=2}^{m+1} \left(\cL_{n+1} - \cL_1(\cQ_0\cL_0\cQ_0)^{-1}\cL_n\right) \rho_{m+1-n} 
&=& 0.
\end{eqnarray}
The superoperator $\cP_1 \cL_2 \cP_1$ is not invertible
since it has a one-dimensional kernel
spanned by $\rho_0$. 
Therefore, we define $\left( \cP_1 \cL_2 \cP_1\right)^{-1}$
as the Moore-Penrose pseudoinverse of $\cP_1 \cL_2 \cP_1$.
Then we have
\begin{eqnarray}
\cP_1 \rho_m &=&
\left(\cP_1 \cL_2 \cP_1\right)^{-1}
\left\{
\left(\cL_1 \left(Q_0 \cL_0 Q_0\right)^{-1}\cL_1 - \cL_2\right)
(\cQ_1 + \cQ_0)\rho_m 
+
\sum_{n=2}^{m+1} \left(\cL_1(\cQ_0\cL_0\cQ_0)^{-1}\cL_n - \cL_{n+1}\right) \rho_{m+1-n} 
\right\}\nonumber\\
&&
+ c\rho_0,
\label{p1rhom}
\end{eqnarray}
where the constant $c$ is determined by the condition
\begin{eqnarray}
{\rm Tr} \rho_m &=& 0\;\;\;\; (m \ge 1).
\end{eqnarray}
Combining Eqs. (\ref{q0rhom}), (\ref{q1rhom}), and (\ref{p1rhom}),
we obtain the $m$th-order solution $\rho_m$.
For example, the first-order solution is given by
\begin{eqnarray}
\rho_1 &=& \cQ_0\rho_1 + \cQ_1 \rho_1 + \cP_1 \rho_1,
\end{eqnarray}
\begin{eqnarray}
\cQ_0 \rho_1 &=& - (\cQ_0 \cL_0 \cQ_0)^{-1}\cL_1 \rho_0,\\
\cQ_1 \rho_1 &=& - (\cQ_1 \cL_1 \cQ_1)^{-1}
\left(\cL_1 \cQ_0 \rho_1 + \cL_2 \rho_0\right),\\
\cP_1 \rho_1 &=&
\left(\cP_1 \cL_2 \cP_1\right)^{-1}
\left\{
\left(\cL_1 \left(\cQ_0 \cL_0 \cQ_0\right)^{-1}\cL_1 - \cL_2\right)
(\cQ_1 + \cQ_0)\rho_1 
+ \left(\cL_1\left(\cQ_0\cL_0\cQ_0\right)^{-1}\cL_2 - \cL_3\right)\rho_0
\right\}\nonumber \\
&&
+ c\rho_0.
\end{eqnarray}

Here, we give some comments about the range of applicability of our
perturbation theory. Our method has two parameters, the
system size and the coupling constant, which compete with each other.
If the system size is large, $\cQ_1\rho_m$
may be very large because 
we have $(\cQ_1\cL_1\cQ_1)^{-1}$ in Eq. (\ref{q1rhom}), whose eigenvalues
may be very large. Then our method could work only for a very small
coupling constant $\lambda$ and might be practically useless for
large systems. However, it is also possible that even when
the higher-order terms of the density matrix are large,
they give only small contributions to local physical quantities
such as local temperature and current. (Note that very different
states, such as a pure state and the Gibbs state, can give
the same expectation values for local observables in macroscopic systems
\cite{sugita2, sugiura}.)
Then our method may be useful even for large systems.

To consider this point, we will give some numerical results for spin
systems in the next section. 
The results show that, for local temperatures, 
the zeroth-order approximation is equally good for the system
sizes $N=2$ and $N=6$.
Therefore, we hope that our method is useful for larger systems.

Another important point to note is that a NESS in a rather small system
may be similar to a part of a macroscopic NESS, at least in some aspects.
Thus, we will be able to obtain some insights on
macroscopic NESSs by studying those in small systems, where
our method will be helpful.

\section{Spin-1/2 Chain} 
\label{spinchain}

In this section, we consider a quantum $N$-spin chain with the Hamiltonian
\begin{eqnarray}
H_{\rm S} &=& H_0 + \lambda H_{\rm I}, 
\end{eqnarray}
\begin{eqnarray}
H_0 &=& h\sum_{l=1}^N\sigma_l^z,\\
 H_{\rm I} &=& \sum_{l=1}^{N-1} \sum_{j,k}\alpha_{j,k}\sigma^j_l
  \sigma^k_{l+1}.
\end{eqnarray}
In this section we set $\hbar = 1$.

\subsection{Basis of $\Ker \cL_0$}
We represent spin up and down states by $|1\rangle$ and $|-1\rangle$,
respectively. Then
$H_0$ has eigenvectors of the form
\begin{eqnarray}
|\{s_l\}\rangle 
&\coloneqq& |s_1\rangle \otimes |s_2\rangle\otimes \dots \otimes |s_N\rangle,
\end{eqnarray}  
where $s_l=\pm 1$, and 
its eigenvalue is $h\sum_l s_l$.
$\Ker \cL_0$ is spanned by elements of the form
\begin{eqnarray}
|\{s_l\}\rangle \langle \{s'_l\}|
\label{basis}
\end{eqnarray}
with 
\begin{eqnarray}
\sum_l s_l = \sum_l s'_l.
\end{eqnarray}
Note that the number of sites with $s_l \ne s'_l$ must be even
to satisfy this condition.

Let us consider some simple cases. If $N=1$, $\Ker\cL_0$ is
spanned by $|1\rangle \langle 1|$ and $|-1\rangle \langle -1|$.
In this case, only diagonal terms appear. Therefore, in this case,
$\Ker\cL_0$ is spanned by $\sigma^0$ and $\sigma^z$,
where $\sigma^0$ is the $2\times 2$ unit matrix.

If $N=2$, $\Ker\cL_0$ is
spanned by six elements,  in which there are
four diagonal elements,
\begin{eqnarray}
|1,1\rangle\langle 1,1|,\;\;\;
|1,-1\rangle\langle 1,-1|,\;\;\;
|-1,1\rangle\langle -1,1|,\;\;\;
|-1,-1\rangle\langle -1,-1|,
\end{eqnarray}
and two off-diagonal elements,
\begin{eqnarray}
|1,-1\rangle \langle -1,1|,\;\;\;
|-1,1\rangle \langle 1,-1|.
\end{eqnarray}
The diagonal terms can be represented
by linear combinations of $\sigma_\alpha\otimes \sigma_\beta$, 
where $\alpha,\beta$ are $0$ or $z$.
The off-diagonal terms are represented as
\begin{eqnarray}
|1,-1\rangle \langle -1,1|
&=& 
\sigma_1^+ \otimes \sigma_2^-
=
\frac{1}{4}
\left\{(\sigma_1^x\otimes \sigma_2^x + \sigma_1^y\otimes \sigma_2^y)
-i (\sigma_1^x\otimes \sigma_2^y - \sigma_1^y\otimes \sigma_2^x)
\right\} \\
|-1,1\rangle \langle 1,-1|
&=&
\sigma_1^-\otimes \sigma_2^+
=
\frac{1}{4}
\left\{(\sigma_1^x\otimes \sigma_2^x + \sigma_1^y\otimes \sigma_2^y)
+ i (\sigma_1^x\otimes \sigma_2^y - \sigma_1^y\otimes \sigma_2^x)
\right\} 
\end{eqnarray}
 where
\begin{eqnarray}
\sigma_l^{\pm} = \frac{1}{2}(\sigma_l^x \pm i \sigma_l^y).
\end{eqnarray}
Therefore, the off-diagonal terms can be represented by
the two terms
$\sigma_1^x\otimes \sigma_2^x + \sigma_1^y\otimes \sigma_2^y$ and
$\sigma_1^x\otimes \sigma_2^y - \sigma_1^y\otimes \sigma_2^y$.

For the $N$-spin case, a basis element in Eq. (\ref{basis}) has an
even number of off-diagonal sites where $s_l \ne s'_l$.
The diagonal parts can be represented by combining 
$\sigma_0$ and $\sigma_z$, and the off-diagonal parts can be
obtained by combining the following two types of correlation
terms:
\begin{eqnarray}
\rho^{\rm A}_{l,m} &\coloneqq& 
\sigma_l^x\otimes \sigma_m^x + \sigma_l^y\otimes \sigma_m^y, 
\label{A}\\
\rho^{\rm B}_{l,m} &\coloneqq& 
\sigma_l^x\otimes \sigma_m^y - \sigma_l^y\otimes \sigma_m^x.
\label{B}
\end{eqnarray}

\subsection{Zeroth-order solution for the two-spin case}
\label{twospin}
Let us consider the two-spin case.
In this case $\Ker \cL_0$ is six dimensional, and we take the
basis as
\begin{eqnarray}
\Ker \cL_{0} &=& \Span \left\{ \sigma^{00}, 
\sigma^{zz}, \sigma^{0z} + \sigma^{z0}, \rho^{\rm A}, \rho^{\rm B}, \rho^{\rm C}
\right\}.
\end{eqnarray}
Here we used the notation
\begin{eqnarray}
\sigma^{jk} &\coloneqq & \sigma_1^j\otimes \sigma_2^k.
\end{eqnarray}
$\rho^{\rm A}$ and $\rho^{\rm B}$ are correlation terms in the form of
Eqs. (\ref{A}) and (\ref{B}), respectively, with $l=1$ and $m=2$.
$\rho^{\rm C}$ is defined as
\begin{eqnarray}
\rho^{\rm C} &\coloneqq& \sigma^{0z} - \sigma^{z0}.
\end{eqnarray}

Then we consider Eq. (\ref{p0l1}). $\cL_1$ is represented by two terms,
\begin{eqnarray}
\cL_1 &=& \cL_{\rm I} + \cL_{\rm E},
\end{eqnarray}
where
\begin{eqnarray}
\cL_{\rm I} \rho &=& -i [H_{\rm I}, \rho],\\
\cL_{\rm E} \rho &=& -i [\overline{H}_{SB}, \rho].
\end{eqnarray}
By straightforward algebra, we obtain
\begin{eqnarray}
\cP_0\cL_{\rm I} \sigma^{00} = \cP_0\cL_{\rm I} \sigma^{zz} 
= \cP_0\cL_{\rm I} (\sigma^{0z} + \sigma^{z0}) = 0,
\end{eqnarray}
\begin{eqnarray}
\cP_0\cL_{\rm I} \rho^{\rm A} &=& 0, \\
\cP_0\cL_{\rm I} \rho^{\rm B} &=& 2\alpha_+ \rho^{\rm C},\\
\cP_0\cL_{\rm I} \rho^{\rm C} &=& -2\alpha_+ \rho^{\rm B}, 
\end{eqnarray}
where $\alpha_+ \coloneqq \alpha_{xx} + \alpha_{yy}$.

We write
\begin{eqnarray}
\overline{H}_{\rm SB} &=& \sum_{j=x,y,z} 
(\alpha_{\rm L}^j \sigma^{j0} + \alpha_{\rm R}^j \sigma^{0j}), 
\end{eqnarray}
which represents the first-order effect of the system-bath coupling
from the left and right heat baths.
Then we have
\begin{eqnarray}
\cP_0\cL_{\rm E} \sigma^{00} = \cP_0\cL_{\rm E} \sigma^{zz} = 
\cP_0\cL_{\rm E} (\sigma^{0z} + \sigma^{z0}) = 0, 
\end{eqnarray}
\begin{eqnarray}
\cP_0 \cL_{\rm E} \rho^{\rm A} &=& -2\Delta\alpha^z \rho^{\rm B},\\
\cP_0 \cL_{\rm E} \rho^{\rm B} &=&  2\Delta\alpha^z \rho^{\rm A},\\
\cP_0 \cL_{\rm E} \rho^{\rm C} &=& 0,
\end{eqnarray}
where $\Delta\alpha^z \coloneqq \alpha_{\rm L}^z - \alpha_{\rm R}^z$.
Therefore, we have
\begin{eqnarray}
\Ker \cP_0\cL_1\cP_0 
&=&
\Span\left\{ \sigma^{00}, \sigma^{zz}, \sigma^{0z} + \sigma^{z0},
\rho^{C'}
\right\},
\end{eqnarray}
where
\begin{eqnarray}
\rho^{C'} &\coloneqq& \rho^{\rm C} - \gamma \rho^{\rm A},
\end{eqnarray}
\begin{eqnarray}
\gamma &\coloneqq& \frac{\alpha_+}{\Delta\alpha_z}.
\end{eqnarray}

Let us consider the local equilibrium form
\begin{eqnarray}
\rho_{\rm LE}(\beta_1)\otimes \rho_{\rm LE}(\beta_2).
\label{le2}
\end{eqnarray}
Here, $\rho_{\rm LE}$ represents a single-spin equilibrium state, 
whose explicit form is
\begin{eqnarray}
\rho_{\rm LE}(\beta_l) = \frac{e^{-\beta_l h\sigma^z}}
{e^{\beta_l h} + e^{-\beta_l h}}
= \frac{1}{2}(\sigma^0 - t_l \sigma^z),
\end{eqnarray}
where $t_l \coloneqq \tanh (\beta_l h)$. Then we apply 
$\cP_0\cL_1$ to Eq. (\ref{le2}). 
Since
\begin{eqnarray}
\rho_{\rm LE}(\beta_1)\otimes \rho_{\rm LE}(\beta_2).
&=&
\frac{1}{4}
\left\{\sigma^{00} + t_1t_2\sigma^{zz} - \frac{t_1 + t_2}{2} 
(\sigma^{0z} + \sigma^{z0})
- \frac{t_2 - t_1}{2}\rho^{\rm C}
\right\},
\end{eqnarray}
we have
\begin{eqnarray}
\cP_0\cL_1 \rho_{\rm LE}(\beta_1)\otimes \rho_{\rm LE}(\beta_2)
&=& 
\frac{t_2 - t_1}{4}\alpha_+ \rho^{\rm B}.
\end{eqnarray}
Therefore, we see that the local equilibrium form in Eq. (\ref{le2})
is not the zeroth-order solution unless $t_1 = t_2$. 
The general form of the solution
of the equation $\cP_0\cL_1 \rho_0 = 0$ is
\begin{eqnarray}
\rho_0 &=& \rho_{\rm LE}(\beta_1)\otimes \rho_{\rm LE}(\beta_2)
+ \frac{t_2 - t_1}{8}\gamma \rho^{\rm A} + g\sigma^{zz},
\label{rho02spin}
\end{eqnarray}
where the three parameters $t_1, t_2$, and $g$ are to be determined from
$\cP_1 \cL_2 \rho_0 = 0$. Note that the correlation term $\rho^{\rm A}$
appears when there is a temperature gradient in the system.

Let us consider the tilted Ising model as an example:
\begin{eqnarray}
H &=& \lambda \sigma^{zz} 
+ 
h\cos\theta(\sigma^{z0} + \sigma^{0z})
+
h\sin\theta(\sigma^{x0} + \sigma^{0x}).
\end{eqnarray}
This model is known to be integrable at $\theta = \pi/2$ 
(transverse Ising model), and
the NESS has a flat temperature profile in this case \cite{saito1}.

By taking the $z$-axis in the direction of the external field,
the Hamiltonian can be transformed as
\begin{eqnarray}
H &=& \lambda\sigma_1^{z'}\otimes \sigma_2^{z'}
+
h (\sigma^{z0} + \sigma^{0z}),
\end{eqnarray}
where
\begin{eqnarray}
\sigma_l^{z'} &\coloneqq& \sigma_l^z\cos\theta - \sigma_l^x\sin\theta.  
\end{eqnarray}
The interaction term can also be written as
\begin{eqnarray}
\sigma_1^{z'}\otimes \sigma_2^{z'}
&=&
\sigma^{zz}\cos^2\theta - 
(\sigma^{zx} + \sigma^{xz}) \cos\theta\sin\theta
+ \sigma^{xx}\sin^2\theta.
\end{eqnarray}
Therefore, we have 
\begin{eqnarray}
\alpha_{zz} &=& \cos^2\theta\\
\alpha_{zx} = \alpha_{xz} &=& -\cos\theta\sin\theta\\ 
\alpha_{xx} &=& \sin^2\theta .
\end{eqnarray}

\begin{figure}[ptb]
\begin{center}
\includegraphics[width=10cm]{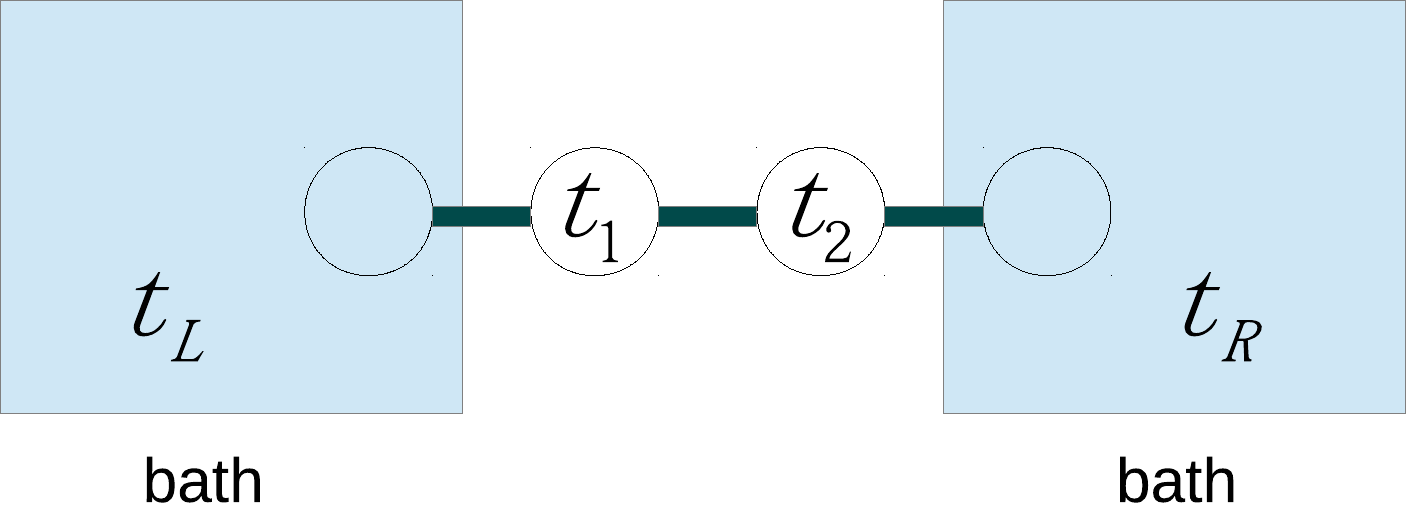}
\caption{Schematic figure of our two-spin model. The system-bath coupling
is assumed to have the same form as the spin-spin coupling. The $t_l$
are related to the local inverse temperatures by $t_l = \tanh(\beta_l h)$.}
\label{2spin_bath_figure}
\end{center}
\end{figure}

We take the spin-bath coupling as having the same form as the spin-spin
coupling (Fig. \ref{2spin_bath_figure}).
This situation may be regarded as similar to that for the two spins in a NESS of a long spin chain.
We also assume that the temperature
difference between the two heat baths is small. Then we obtain
\begin{eqnarray}
t_1 &=& \frac{t_{\rm L} + t_{\rm R}}{2} + \frac{K}{2}(t_{\rm L} - t_{\rm
 R}),
 \label{2spin_temp1}\\
t_2 &=& \frac{t_{\rm L} + t_{\rm R}}{2} - \frac{K}{2}(t_{\rm L} - t_{\rm
 R}),
 \label{2spin_temp2}\\
g &=& - \frac{K(1-K)}{16}(t_{\rm L} - t_{\rm R})^2,
\end{eqnarray} 
where $t_l = \tanh(\beta_l h)\;\; (l = {\rm L,R})$, and $\beta_{\rm L}$ 
($\beta_{\rm R}$)
is the inverse temperature 
for the left (right) heat bath.
$K$ is defined as
\begin{eqnarray}
K &\coloneqq & 
\frac{c(t_{\rm L} - t_{\rm R})^2\cos^4\theta}
{c(t_{\rm L} - t_{\rm R})^2\cos^4\theta 
+ \sin^2\theta(2d\cos^2\theta + c\sin^2\theta)},
\end{eqnarray}
where $c$ and $d$ are non-negative constants.
(See Appendix\ref{detail} for details of the calculation.)
We see that a nonzero temperature gradient is formed unless
$\alpha_{zz} = \cos^2\theta = 0$, which corresponds to the
transverse Ising model.

\begin{figure}[htbp]
    \begin{minipage}{0.48\textwidth}
  \begin{center}
   \includegraphics[width=1.3\textwidth]{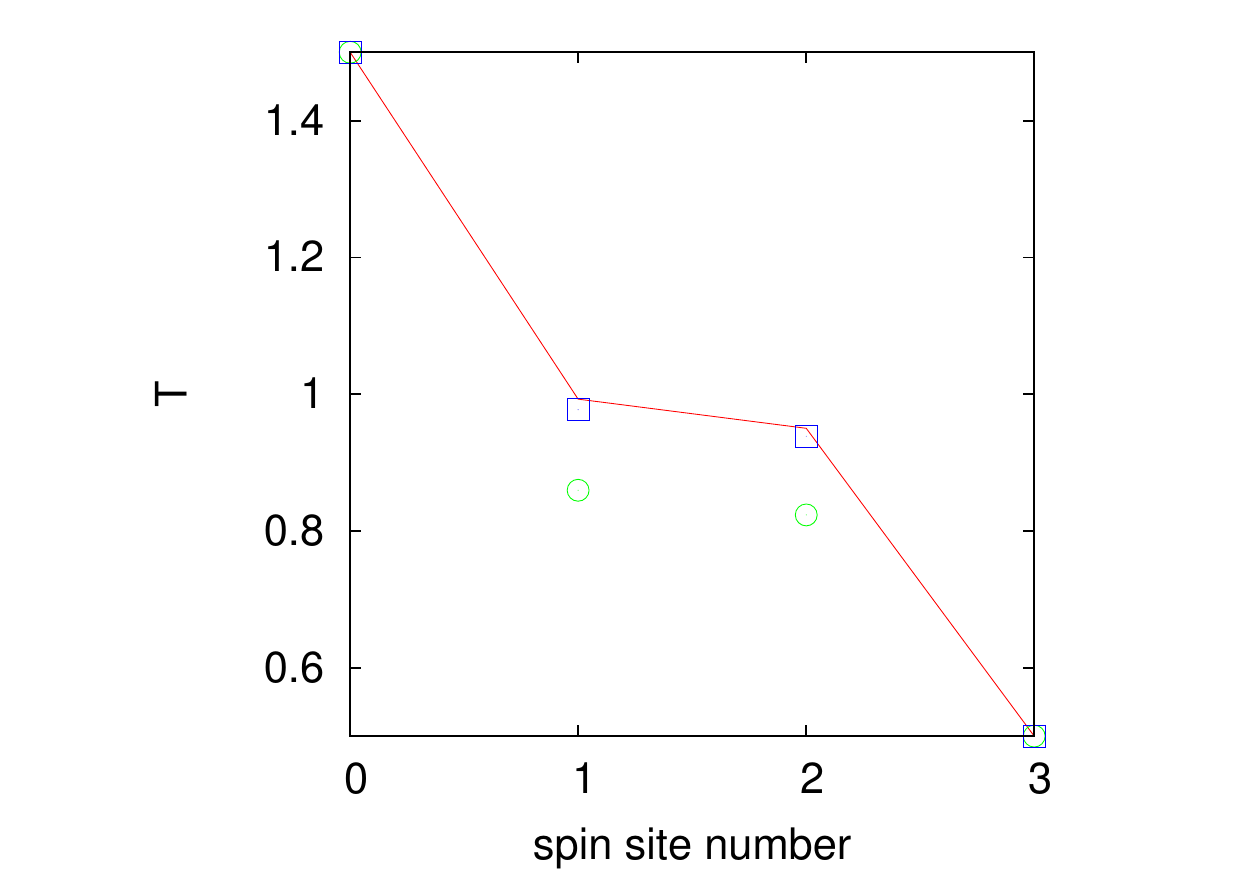}
  \end{center}
 \end{minipage}
 \begin{minipage}{0.48\textwidth}
  \begin{center}
   \includegraphics[width=1.3\textwidth]{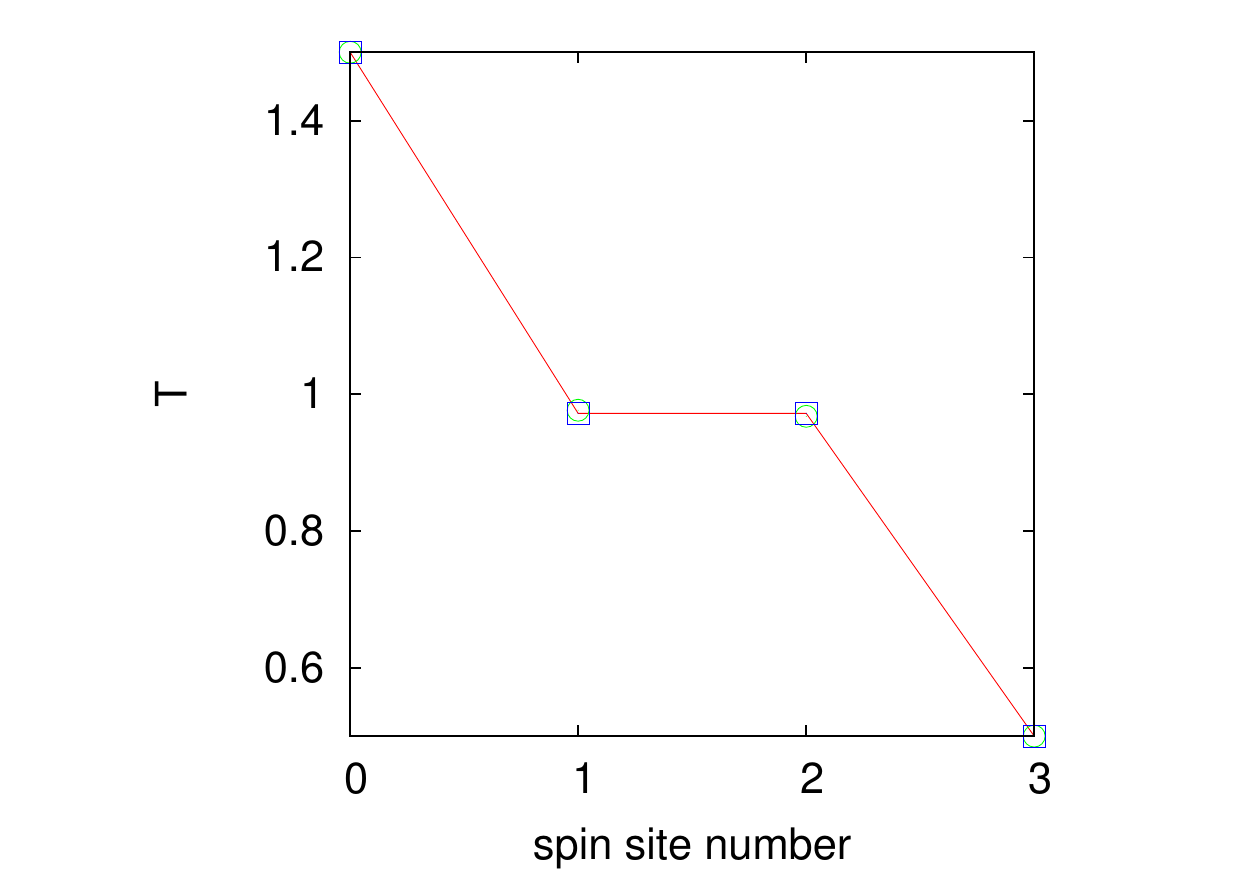}
  \end{center}
 \end{minipage}
  \caption{Temperature profiles for the two-spin tilted Ising model.
  The left and right figures show the results for $\theta = \pi/4$
  and $\theta = \pi/2$ (transverse Ising model), respectively.
  The heat bath temperatures are $T_L=1.5$ and $T_R=0.5$, which
  are shown at the zeroth and third sites in the figure, respectively.
  The solid line shows the analytical results in Eqs. (\ref{2spin_temp1}) and
  (\ref{2spin_temp2}). The circles and squares are numerical
  results for $\lambda = 0.1$ and $0.01$, respectively.}
  \label{2spin_figure}
 \end{figure}

Figure \ref{2spin_figure} shows the temperature profiles for the
tilted Ising model with $N=2$. The numerical results are obtained
by diagonalizing the Liouvillian in Eq. (\ref{liouvillian}).
We set $h=1$ and take the heat bath parameters
as $c_L = c_R = c_L = c_R = 1$ in Eqs. (\ref{c's}) and (\ref{d's}).
Then Eqs. (\ref{2spin_temp1}) and (\ref{2spin_temp2}) should hold
 exactly in the limit $\lambda\rightarrow 0$. In the case with $\theta=\pi/4$,
 the numerical results are in good
 agreement with the analytical results for $\lambda=0.01$ with
 relative errors of order $10^{-2}$. For the case with $\theta=\pi/2$
 (transverse Ising model), where the temperature gradient vanishes,
 the convergence to the weak-coupling limit is much faster.

\subsection{$N$-spin case}

Next we consider $N$-spin systems. Let us first 
consider the product of local equilibrium states:
\begin{eqnarray}
\rho_{\rm PLE} &\coloneqq & \bigotimes_{m=1}^N \rho_{\rm LE}(\beta_m).
\end{eqnarray}
We write $\cL_1$ as
\begin{eqnarray}
\cL_1 &=& \sum_{l=1}^{N-1}\cL_{{\rm I}(l,l+1)} + \cL_{{\rm E}(1)} + \cL_{{\rm E}(N)},
\end{eqnarray}
where $\cL_{{\rm I}(l,l+1)}$ represents a local interaction,
\begin{eqnarray}
\cL_{{\rm I}(l,l+1)}\rho &=& -i [H_{{\rm I}(l,l+1)}, \rho],
\end{eqnarray}
\begin{eqnarray}
H_{{\rm I}(l,l+1)} &\coloneqq & 
\sum_{jk}\alpha_{jk}\sigma^j_{l}\otimes \sigma^k_{l+1},
\end{eqnarray}
and $\cL_{{\rm E}(l)}$ represents the first-order contribution from a heat bath,
\begin{eqnarray}
\cL_{{\rm E}(l)} \rho &=& -i [H_{{\rm E}(l)}, \rho],
\end{eqnarray}
\begin{eqnarray}
H_{{\rm E}(l)} &\coloneqq& \alpha_l^x \sigma_l^x + \alpha_l^y\sigma_l^y + \alpha_l^z\sigma_l^z.
\end{eqnarray}
Then we have
\begin{eqnarray}
\cP_0 \cL_{\rm I(l,l+1)}\rho_{\rm PLE} &=& 
\frac{t_{l+1}-t_l}{4}\alpha_+\rho^{\rm B}_{l,l+1}\bigotimes_{m(\ne l,l+1)}
\rho_{\rm LE}(\beta_m)
\label{PLEcontribution}
\end{eqnarray}
and
\begin{eqnarray}
\cP_0 \cL_{{\rm E}(l)}\rho_{\rm PLE} &=& 0.
\end{eqnarray}
Since $\cP_0\cL_1\rho_{\rm PLE}\ne 0$,
$\rho_{\rm PLE}$ is not the zeroth-order solution. 
Therefore we have to add some correction terms to $\rho_{\rm PLE}$ to obtain
$\rho_0$. To cancel the two-body correlated term in Eq. (\ref{PLEcontribution}),
we have to add three-body correlated terms, 
but if we apply $\cP_0\cL_1$ to
them, slightly different residual three-body terms appear.
Then we have to add four-body correlated terms to cancel the
three-body residual terms, which give new four-body residual terms, and so on. Thus, we obtain a BBGKY (Bogoliubov-Born-Green-Kirkwood-Yvon)-like hierarchy 
of equations, which is difficult to solve. 

In this paper, we do not try to obtain $\rho_0$ explicitly.
Instead, we consider local projections of the equation
$\cP_0\cL_1\rho_0 = 0$. A one-site projection $\cP_{(l)}$ traces out
all spins except for that on the $l$th site. It gives no
information, however, since $\rho_{\rm PLE}$ already satisfies
\begin{eqnarray}
\cP_{(l)}\cP_0 \cL_1 \rho_{\rm PLE} &=& 0.
\end{eqnarray}

Next we consider a two-site projection
$\cP_{(l,l+1)}$, which traces out all spins except for those at
the $l$th and $(l+1)$th sites.
Since 
\begin{eqnarray}
\cP_{(l,l+1)}\cP_0 \cL_{{\rm I}(m,m+1)} = 0
\end{eqnarray}
if $l+1<m$ or $m+1<l$, only three interaction terms,
$\cL_{{\rm I}(l-1,l)}$, $\cL_{{\rm I}(l,l+1)}$, and $\cL_{{\rm I}(l+1,l+2)}$ matter
when we consider $\cP_{(l,l+1)}$. Hence, the equation to solve is
\begin{eqnarray}
\cP_{(l,l+1)} \cP_0
\left( \cL_{{\rm I}(l-1,l)} + \cL_{{\rm I}(l,l+1)} +\cL_{{\rm I}(l+1,l+2)}\right) \rho_0
&=& 0.
\label{twosite}
\end{eqnarray}
We write
\begin{eqnarray}
\rho_0 &=& \rho_{\rm PLE} + \rho_{a}.
\label{add}
\end{eqnarray}
Without loss of generality, we can assume that
the additional term $\rho_{a}$ does not change the local 
state: 
\begin{eqnarray}
\cP_{l}\rho_0 &=& \rho_{\rm LE}(\beta_l).
\end{eqnarray}
Then the general form of $\rho_a$ that is relevant to Eq. (\ref{twosite})
is
\begin{eqnarray}
\rho_a &=& 
c^{\rm A}_{\overline{l-1},l,l+1}\sigma_{l-1}^z \rho^{\rm A}_{l,l+1} +
c^{\rm B}_{\overline{l-1},l,l+1}\sigma_{l-1}^z \rho^{\rm B}_{l,l+1} +
c^{\rm A}_{l-1,\overline{l},l+1}\sigma_{l}^z \rho^{\rm A}_{l-1,l+1} +
c^{\rm B}_{l-1,\overline{l},l+1}\sigma_{l}^z \rho^{\rm B}_{l-1,l+1} \nonumber\\
&&
+
c^{\rm A}_{l,\overline{l+1},l+2}\sigma_{l+1}^z \rho^{\rm A}_{l,l+2} +
c^{\rm B}_{l,\overline{l+1},l+2}\sigma_{l+1}^z \rho^{\rm B}_{l,l+2} +
c^{\rm A}_{l,l+1,\overline{l+2}}\rho^{\rm A}_{l,l+1} \sigma_{l+2}^z +
c^{\rm B}_{l,l+1,\overline{l+2}}\rho^{\rm B}_{l,l+1} \sigma_{l+2}^z\nonumber\\
&&
+
c^{\rm B}_{l,l+1}\rho^{\rm B}_{l,l+1}.
\label{general}
\end{eqnarray}
In the coefficients $c^{\alpha}_{p,q,r}$, 
the overline attached to a suffix shows the site where
$\sigma^z$ is.
We assume that, at a site whose state is unspecified,
the state is $\sigma^0/2$, whose trace is unity. For example,
\begin{eqnarray}
\sigma_{l-1}^z \rho^{\rm A}_{l,l+1} &\coloneqq& 
\sigma_{l-1}^z \rho^{\rm A}_{l,l+1} \bigotimes_{m(\ne l-1,l,l+1)}\frac{\sigma_m^0}{2}.
\end{eqnarray}
Substituting Eq. (\ref{general}) into Eqs. (\ref{twosite}) and (\ref{add}), 
we obtain
\begin{eqnarray}
&& \left\{
\frac{t_{l+1}-t_l}{4}\alpha_+ +
4\alpha_{zz}(-c^{\rm A}_{\overline{l-1},l,l+1} + c^{\rm A}_{l,l+1,\overline{l+2}})
+ 2\alpha_+ (c^{\rm A}_{l-1,\overline{l},l+1} - c^{\rm A}_{l,\overline{l+1},l+2})\right\}
\rho^{\rm B}_{l,l+1} \nonumber\\
 &&
  - 2\left\{2\alpha_{zz}(-c^{\rm B}_{\overline{l-1},l,l+1} + c^{\rm B}_{l,l+1,\overline{l+2}})
+ \alpha_+ (c^{\rm B}_{l-1,\overline{l},l+1} - c^{\rm B}_{l,\overline{l+1},l+2})
\right\}
\rho^{\rm A}_{l,l+1} \nonumber\\
&&
+ 2\alpha_+ c^{\rm B}_{l,l+1}\rho_{l,l+1}^C = 0,
\end{eqnarray}
where $\rho^{\rm C}_{l,m} \coloneqq 
\sigma^0_l \sigma^z_m - \sigma^z_l \sigma^0_m$. Hence,
\begin{eqnarray}
(t_{l+1}-t_l)\alpha_+ +
16\alpha_{zz}(-c^{\rm A}_{\overline{l-1},l,l+1} + c^{\rm A}_{l,l+1,\overline{l+2}})
+ 8\alpha_+ (c^{\rm A}_{l-1,\overline{l},l+1} - c^{\rm A}_{l,\overline{l+1},l+2}) &=& 0,\\
2\alpha_{zz}(-c^{\rm B}_{\overline{l-1},l,l+1} + c^{\rm B}_{l,l+1,\overline{l+2}})
+ \alpha_+ (c^{\rm B}_{l-1,\overline{l},l+1} - c^{\rm B}_{l,\overline{l+1},l+2}) &=& 0,\\
c^{\rm B}_{l,l+1} &=& 0.
\label{spin_current}
\end{eqnarray}

At the edges of the system, we have slightly different equations. For example,
at the left edge ($l=1$), we have
\begin{eqnarray}
\cP_{(1,2)} \cP_0 (\cL_{{\rm E}(1)} + \cL_{{\rm I}(1,2)} + \cL_{{\rm I}(2,3)})\rho_0 &=& 0.
\label{eq_edge}
\end{eqnarray} 
The general form of the relevant additional term is
\begin{eqnarray}
\rho_a &=& 
c^{\rm A}_{1,\overline{2},3}\sigma_{2}^z \rho^{\rm A}_{1,3} +
c^{\rm B}_{1,\overline{2},3}\sigma_{2}^z \rho^{\rm B}_{1,3} +
c^{\rm A}_{1,2,\overline{3}}\rho^{\rm A}_{1,2} \sigma_{3}^z +
c^{\rm B}_{1,2,\overline{3}}\rho^{\rm B}_{1,2} \sigma_{3}^z +
c^{\rm A}_{1,2}\rho^{\rm A}_{1,2} + 
c^{\rm B}_{1,2}\rho^{\rm B}_{1,2}.
\label{add_edge}
\end{eqnarray}
Substituting Eq. (\ref{add_edge}) into Eq. (\ref{eq_edge}), we obtain
\begin{eqnarray}
&& \left\{
\frac{t_{2}-t_1}{4}\alpha_+ +
4\alpha_{zz}c^{\rm A}_{1,2,\overline{3}}
- 2\alpha_+ c^{\rm A}_{1,\overline{2},3}
- 2 \alpha_1^z c^{\rm A}_{1,2}
\right\}
\rho^{\rm B}_{1,2} \nonumber\\
 &&
  - 2\left\{2\alpha_{zz}c^{\rm B}_{1,2,\overline{3}}
- \alpha_+ c^{\rm B}_{1,\overline{2},3} - \alpha_1^z c^{\rm B}_{1,2}
\right\}
\rho^{\rm A}_{1,2} \nonumber\\
&&
+ 2\alpha_+ c^{\rm B}_{1,2}\rho_{1,2}^C = 0,
\end{eqnarray}
Hence,
\begin{eqnarray}
(t_{2}-t_1)\alpha_+ +
16\alpha_{zz}c^{\rm A}_{1,2,\overline{3}}
- 8\alpha_+ c^{\rm A}_{1,\overline{2},3}
- 8 \alpha_1^z c^{\rm A}_{1,2} &=& 0, \\
2\alpha_{zz}c^{\rm B}_{1,2,\overline{3}}
- \alpha_+ c^{\rm B}_{1,\overline{2},3} &=& 0, \\
c^{\rm B}_{1,2} &=& 0.
\end{eqnarray}
In the same way, we obtain the following three equations at the
right edge:
\begin{eqnarray}
(t_{N}-t_{N-1})\alpha_+ 
- 16\alpha_{zz}c^{\rm A}_{\overline{N-2},N-1,N}
+ 8\alpha_+ c^{\rm A}_{N-2,\overline{N-1},N}
+ 8 \alpha_{N}^z c^{\rm A}_{N-1,N} &=& 0, \\
2\alpha_{zz}c^{\rm B}_{\overline{N-1},N-1,N}
- \alpha_+ c^{\rm B}_{N-2,\overline{N-1},N} &=& 0, \\
c^{\rm B}_{N-1,N} &=& 0.
\end{eqnarray}

Thus, we have obtained the following set of equations related to the temperature
gradient:
\begin{eqnarray}
t_2 - t_1 &=& \frac{8}{\alpha_+}(- 2 \alpha_{zz}c^{\rm A}_{1,2,\overline{3}}
+ \alpha_+ c^{\rm A}_{1,\overline{2},3}
+ \alpha_1^z c^{\rm A}_{1,2}), \\
t_3 - t_2 &=& 
 \frac{8}{\alpha_+}
 \left\{2\alpha_{zz}(c^{\rm A}_{\overline{1},2,3} - c^{\rm A}_{2,3,\overline{4}})
+ \alpha_+(-c^{\rm A}_{1,\overline{2},3} + c^{\rm A}_{2,\overline{3},4})\right\},\\
 t_4 - t_3 &=&
  \frac{8}{\alpha_+}
\left\{2\alpha_{zz}(c^{\rm A}_{\overline{2},3,4} - c^{\rm A}_{3,4,\overline{5}})
+ \alpha_+(-c^{\rm A}_{2,\overline{3},4} + c^{\rm A}_{3,\overline{4},5})\right\},\\
&\vdots & \\
 t_{N-1} - t_{N-2} &=&
  \frac{8}{\alpha_+}
\left\{2\alpha_{zz}(c^{\rm A}_{\overline{N-3},N-2,N-1} - c^{\rm A}_{N-2,N-1,\overline{N}})
+ \alpha_+(-c^{\rm A}_{N-3,\overline{N-2},N-1} + c^{\rm A}_{N-2,\overline{N-1},N})\right\},\\
 t_N - t_{N-1} &=&
  \frac{8}{\alpha_+}
(2\alpha_{zz}c^{\rm A}_{\overline{N-2},N-1,N}
- \alpha_+ c^{\rm A}_{N-2,\overline{N-1},N}
- \alpha_N^z c^{\rm A}_{N-1,N}). 
\end{eqnarray}
Summing these equations, all terms with $\alpha_+$ vanish and
we obtain
\begin{eqnarray}
 t_N - t_1 &=& \frac{8}{\alpha_+}
  \left\{2\alpha_{zz}\sum_{l=2}^{N-1}
\left(c^{\rm A}_{\overline{l-1},l,l+1} - c^{\rm A}_{l-1,l,\overline{l+1}}\right)
+ \alpha_1^zc_{1,2}^{\rm A} - \alpha_N^zc_{N-1,N}^{\rm A}\right\}.
\end{eqnarray}
If we assume
that the system-bath coupling has the same form
as the spin-spin coupling as in Sect. \ref{twospin}, we have $\alpha_1^z = -\alpha_{zz}t_{\rm L}$ and 
$\alpha_N^z = -\alpha_{zz}t_{\rm R}$. Replacing the coefficients
with the expectation values of the correlations, we obtain
\begin{eqnarray}
 t_N - t_1
  &=& \frac{\alpha_{zz}}{\alpha_+}
  \left\{
   \sum_{l=2}^{N-1}
   \left(\langle \sigma_{l-1}^z\rho^{\rm A}_{l,l+1}\rangle
    - \langle \rho^{\rm A}_{l-1,l}\sigma^z_{l+1}\rangle\right)
   - t_{\rm L}\langle\rho^{\rm A}_{1,2}\rangle
   + \langle \rho^{\rm A}_{N-1,N}\rangle t_{\rm R}\right\}.
  \label{main}
\end{eqnarray}
This is our main result. The left-hand side (LHS) is related 
to the temperature difference
between the two ends of the chain.
Here, we see that $\alpha_{zz}\ne 0$ is necessary
to form a nonzero temperature gradient. We can also see that
the three-body correlations 
\begin{eqnarray}
\langle \sigma_{l-1}^z \rho^{\rm A}_{l,l+1}\rangle &=&
\langle \sigma_{l-1}^z(\sigma^x_l\sigma^x_{l+1} + \sigma^y_l\sigma^y_{l+1}) \rangle
\end{eqnarray}
and
\begin{eqnarray}
\langle \rho^{\rm A}_{l-1,l}\sigma_{l+1}\rangle &=&
\langle (\sigma^x_{l-1}\sigma^x_{l} + \sigma^y_{l-1}\sigma^y_{l})
\sigma_{l+1}^z \rangle
\end{eqnarray}
play essential roles in forming the temperature gradient.
Note that the two-body correlation terms related to the heat baths
also have similar forms since $t_l = -\langle \sigma_l\rangle$. 

\begin{figure}[htbp]
 \begin{minipage}{0.48\hsize}
  \begin{center}
   \includegraphics[width=1.3\textwidth]{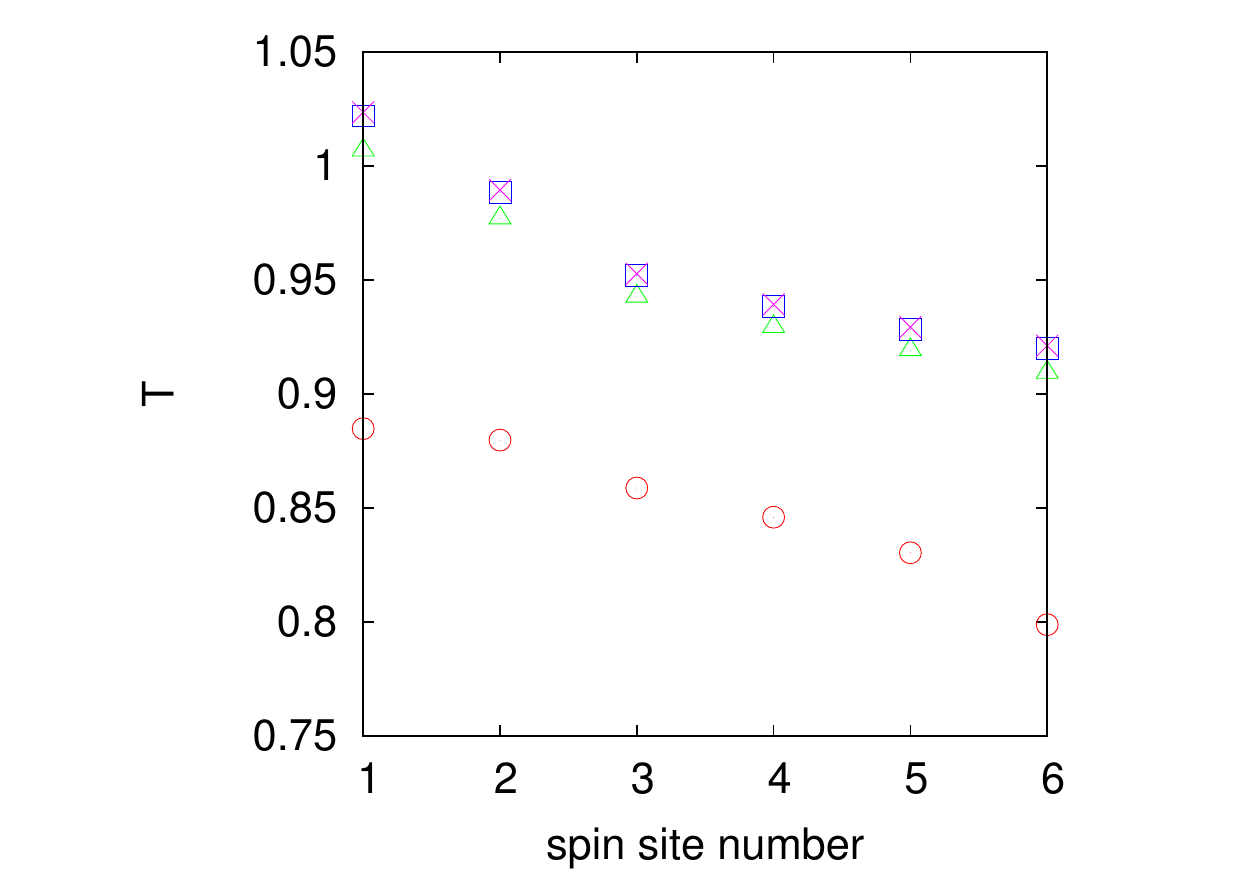}
  \end{center}
 \end{minipage}
 \begin{minipage}{0.48\hsize}
  \begin{center}
   \includegraphics[width=1.3\textwidth]{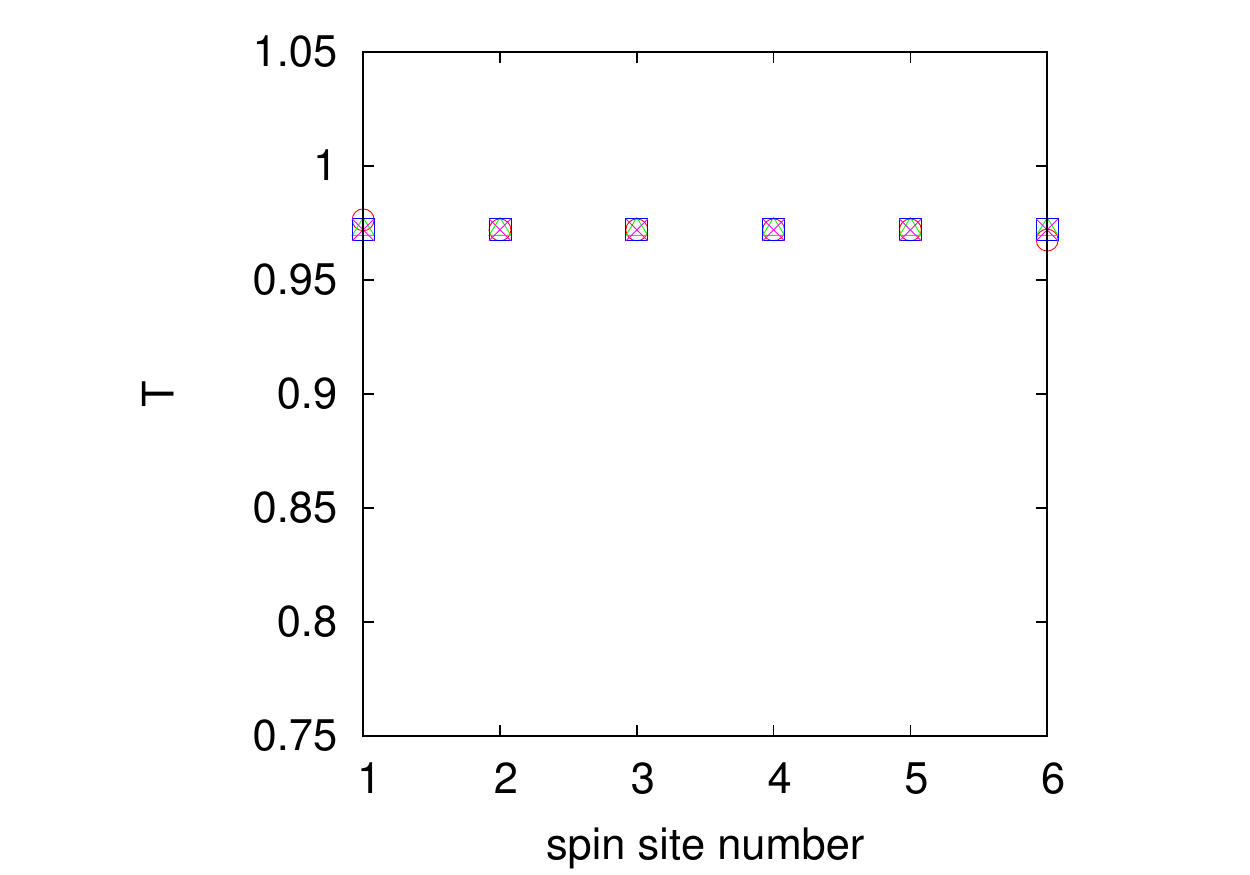}
  \end{center}
 \end{minipage}
 \caption{Temperature profiles for the tilted Ising model with $N=6$.
 The left and right figures
 correspond to $\theta=\pi/4$ and $\theta=\pi/2$, respectively.
 The circles, triangles, squares, and crosses show the numerical results for
 $\lambda = 10^{-1}, 10^{-2}, 10^{-3}$, and $10^{-4}$, respectively.
 The other parameters, which are not written explicitly here, are the
 same as those for the two-spin case in Fig. \ref{2spin_figure}.}
  \label{6spin_figure}
\end{figure}

In Fig. \ref{6spin_figure} we show numerically obtained temperature
profiles for the tilted Ising model with $N=6$.
We see  that the temperature gradient is nonzero for
$\theta=\pi/4$, where $\alpha_{zz}\ne 0$, while
it vanishes for $\theta=\pi/2$,
where $\alpha_{zz}= 0$, as expected from Eq. (\ref{main}).
In the case with $\theta=\pi/4$,
the numerical results agree with the weak-coupling
limit for $\lambda=0.01$, with relative errors of order $10^{-2}$.
The convergence is as fast as for the two-spin case. 
For $\theta=\pi/2$, the convergence is much faster, which
is also the same as in the two-spin case.

\begin{figure}[htbp]
 \begin{center}
  \includegraphics[width=0.7\textwidth]{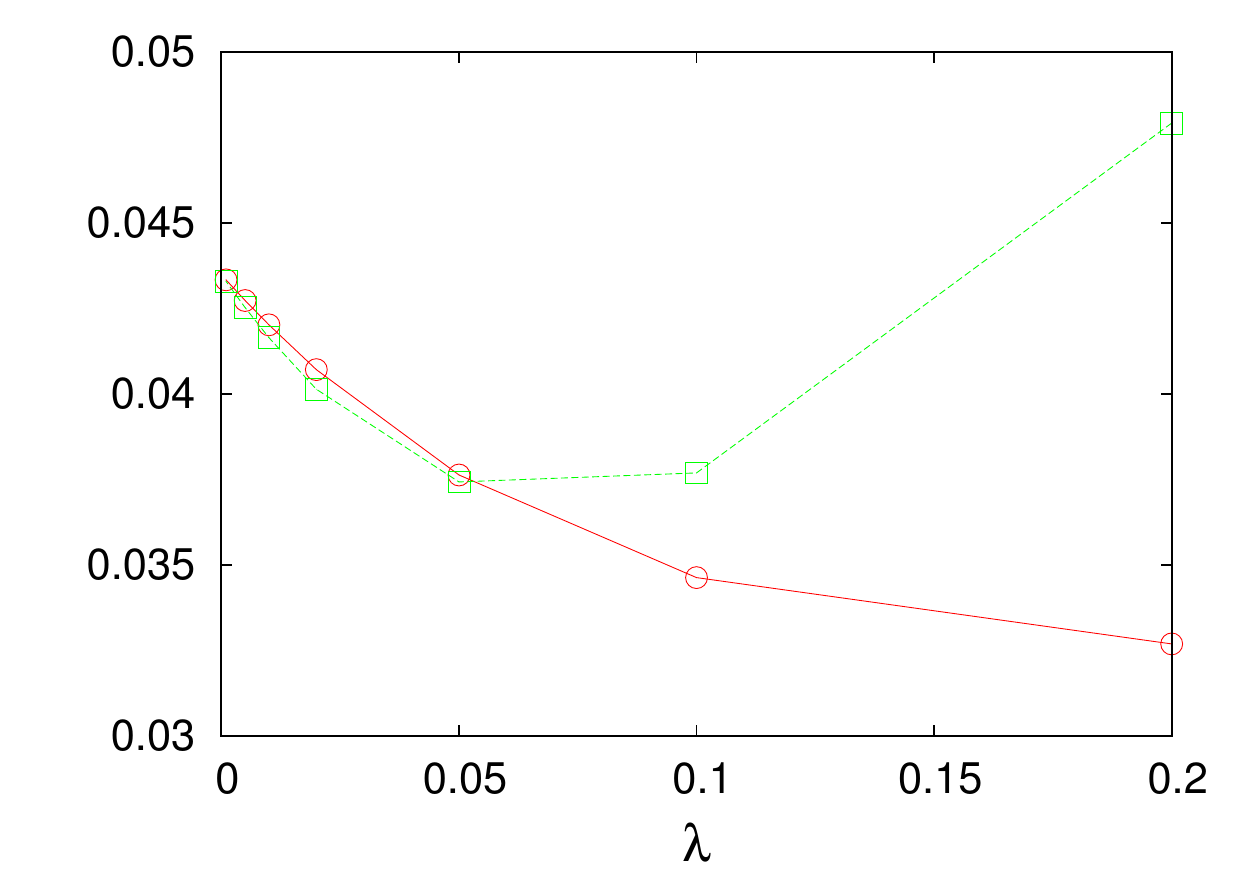}
  \caption{Numerical results
  for the LHS (circles) and RHS (squares) of Eq. (\ref{main}).
  The coupling parameter is taken as
  $\lambda=0.001,0.005, 0.01, 0.02, 0.05, 0.1$,
  and $0.2$.}
 \label{corr_fig}
 \end{center}
\end{figure}

In Fig. \ref{corr_fig} we show numerical results for the LHS and the 
right-hand side (RHS) of
 Eq. (\ref{main}). We see that both sides converge to the same value
 in the weak-coupling limit, although the convergence is slightly slower
 than that for the temperature.

\section{Summary}
\label{summary}
We considered NESSs in weakly coupled one-dimensional quantum spin-1/2
chains. The spin-spin coupling and system-bath coupling were
assumed to be of the same order. We derived the QME
for this case, in which the heat bath superoperator becomes local. We
developed a perturbative expansion for the stationary solution of
the QME. For the two-spin case, we gave an explicit 
example in
which the zeroth-order solution has a nonzero temperature gradient.
For the general $N$-spin case, we showed that $\alpha_{zz}\ne 0$
is necessary to obtain a nonzero temperature gradient. We also showed
that some three-body correlations are essential for the formation of
a temperature gradient.

\begin{acknowledgments}
The authors thank Akira Terai, Katsuhiro Nakamura, and Shogo Tanimura
 for helpful discussions and comments.
\end{acknowledgments}


\appendix

\section{Proof of Eq. (\ref{mendou})}
\label{proof_mendou}
Let us denote a matrix element of 
$H_1$ as
\begin{eqnarray}
d_{n,k,m,l} \coloneqq \langle E_{n,k}|H_1 |E_{m,l}\rangle. 
\end{eqnarray} 
It satisfies
\begin{eqnarray}
d_{n,k,m,l} &=& d_{m,l,n,k}^*.
\end{eqnarray}
We apply $\cL_1 \left(Q_0 \cL_0 Q_0 \right)^{-1} \cL_1 \cP_1$
to a basis element $|E_{n,k}\rangle \langle E_{n,k}|$ in $\tIm \cP_1$.
Then we obtain
\begin{eqnarray}
&&
\cL_1 \left(\cQ_0\cL_0\cQ_0\right)^{-1}\cL_1 |E_{n,k}\rangle \langle E_{n,k}|
\nonumber\\
&=&
\frac{1}{i\hbar}
\sum_{p,q}\sum_{m(\ne n),l} 
\frac{1}{E_m - E_n}
\left\{
|E_{p,q}\rangle d_{p,q,m,l} d_{m,l,n,k} \langle E_{n,k}| 
-
|E_{m,l}\rangle d_{m,l,n,k} d_{n,k,p,q} \langle E_{p,q}|
\right.\nonumber\\
&&
\left.
+
|E_{p,q}\rangle d_{p,q,n,k}d_{n,k,m,l}\langle E_{m,l}|
-
|E_{n,k}\rangle d_{n,k,m,l}d_{m,l,p,q}\langle E_{p,q}|
\right\}.
\end{eqnarray}
Applying $\cP_1$ to the above equation, we have
\begin{eqnarray}
&&
\cP_1\cL_1 \left(\cQ_0\cL_0\cQ_0\right)^{-1}\cL_1 |E_{n,k}\rangle \langle E_{n,k}|
\nonumber\\
&=&
\frac{1}{i\hbar}
\sum_{m(\ne n),l} 
\frac{1}{E_m - E_n}
\left\{
|E_{n,k}\rangle |d_{m,l,n,k}|^2 \langle E_{n,k}| 
-
|E_{m,l}\rangle |d_{m,l,n,k}|^2\langle E_{m,l}|
\right.\nonumber\\
&&
\left.
+
|E_{m,l}\rangle |d_{m,l,n,k}|^2\langle E_{m,l}|
-
|E_{n,k}\rangle |d_{m,l,n,k}|^2\langle E_{n,k}|
\right\} \\
&=&
0.
\end{eqnarray}
Hence, Eq. (\ref{mendou}) is proven.

\section{Two-spin tilted Ising model}
\label{detail}

In this appendix, we show the details of the calculation
of the zeroth-order solution for the two-spin tilted Ising model.

First we consider a single bath and assume spin-bath 
coupling of the form $\sigma^{z'}\otimes \sigma^{z'}$,
where
\begin{eqnarray}
\sigma^{z'} &=& \sigma^z \cos\theta - \sigma^x \sin\theta. 
\end{eqnarray}
Then the heat bath superoperator becomes
\begin{eqnarray}
\cD \rho = -\left([\sigma^{z'}, S\rho] + h.c.\right),
\end{eqnarray}
where
\begin{eqnarray}
S &=& 
\left(
\begin{array}{cc}
\Xi_\beta(0)\cos\theta & -\Xi_\beta(2h)\sin\theta \\
-\Xi_\beta (-2h)\sin\theta & - \Xi_\beta(0)\cos\theta 
\end{array}
\right),
\end{eqnarray}
\begin{eqnarray}
\Xi_\beta (\omega) &=& \int_0^\infty dt e^{-i\omega t}\Phi_\beta(t),
\end{eqnarray}
\begin{eqnarray}
\Phi_\beta (t) &=& \langle \Delta \sigma^{z'}(t)\Delta \sigma^{z'}\rangle_\beta.
\end{eqnarray}
Here, $\langle \cdot \rangle_\beta$ represents the equilibrium average
in the heat bath at the inverse temperature $\beta$. We removed suffices
$j,l$ from $\Xi$ and $\Phi$ since the system-bath coupling is represented
by a single term, and use the subscript 
$\beta$ to show the inverse temperature of the heat bath
explicitly. Note that we take $\hbar = 1$ here.

$\Xi_\beta$ can be written as
\begin{eqnarray}
\Xi_\beta (\omega) &=& \frac{1}{2}
\left\{\tilde{\Phi}_\beta(\omega) + i\Psi_\beta(\omega)\right\},
\end{eqnarray}
where the real part is represented by the Fourier transform of $\Phi_\beta(t)$:
\begin{eqnarray}
\tilde\Phi_\beta(\omega) &=& \int_{-\infty}^\infty dt e^{-i\omega t}\Phi_\beta (t).
\end{eqnarray}
It satisfies the KMS (Kubo-Martin-Schwinger) condition \cite{kubo}
\begin{eqnarray}
\tilde{\Phi}_\beta(-\omega) &=& \tilde{\Phi}_\beta(\omega)e^{\beta\omega}
\end{eqnarray}
and $\tilde{\Phi}(\omega)\ge 0$.
We ignore the imaginary part $\Psi_\beta$ because usually it does not
give a significant contribution.\cite{saito2} 

Then
\begin{eqnarray}
S &=& \frac{1}{2}\left\{b\sigma_z - a(\sigma_x - it\sigma_y)\right\},
\end{eqnarray}
where
\begin{eqnarray}
a &=& 
\frac{\tilde{\Phi}_\beta (2h) + \tilde{\Phi}_\beta(-2h)}{2}\sin\theta,\\
b &=& \tilde{\Phi}_\beta (0)\cos\theta, \\
t &=& \tanh(\beta h).
\end{eqnarray}

Thus, we have
\begin{eqnarray}
\cD \sigma_0 &=& -2at(\sigma_x\cos\theta + \sigma_z\sin\theta), \\
\cD \sigma_x &=& -2b(\sigma_x\cos\theta - \sigma_z\sin\theta), \\
\cD \sigma_y &=& -2(b\cos\theta + a\sin\theta)\sigma_y, \\
\cD \sigma_z &=& -2a(\sigma_x\cos\theta + \sigma_z\sin\theta).
\end{eqnarray}

Next we apply $\cP_1 \cL_2 = \cP_1 (\cD_{\rm L} + \cD_{\rm R})$ to Eq. (\ref{rho02spin}).
Then we obtain
\begin{eqnarray}
&&
\cP_1 \cL_2 \rho_0\\
&=&
- \frac{\sin\theta}{2}\left\{
a_L(t_1 - t_{\rm L})t_2 + a_R(t_2 - t_{\rm R})t_1
+ 4g (a_L + a_R)\right\}\sigma^{zz} \nonumber\\
&&
+ \frac{\sin\theta}{4}\left\{ a_L(t_1-t_{\rm L}) + a_R(t_2 - t_{\rm R})
\right\}(\sigma^{0z} + \sigma^{z0})\nonumber\\
&&
+ \frac{1}{8(1+\gamma^2)}
\left[2\sin\theta\left\{a_R(t_2-t_{\rm R}) - a_L(t_1-t_{\rm
		  L})\right\}\right. \nonumber \\
&&
\left.
+ \gamma^2(t_2-t_1)
\left\{2(b_L + b_R)\cos\theta + (a_L + a_R)\sin\theta\right\}
\right]\rho^{C'},
\end{eqnarray}
with obvious notations. Since in this case
\begin{eqnarray}
\overline{H}_{SB} &=& 
\langle \sigma^{z'}\rangle_{\beta_L}\sigma_1^{z'}\otimes\sigma_2^0
+
\sigma_1^0\otimes\sigma_2^{z'}\langle \sigma^{z'}\rangle_{\beta_R}\\
&=&
- \cos\theta\left(t_{\rm L} \sigma_1^{z'}\otimes\sigma_2^0
+ t_{\rm R} \sigma_1^0\otimes\sigma_2^{z'}\right),
\end{eqnarray}
we have
\begin{eqnarray}
\Delta\alpha_z = - (t_{\rm L} - t_{\rm R})\cos^2\theta
\end{eqnarray}
and
\begin{eqnarray}
\gamma = \frac{\alpha_{xx} + \alpha_{yy}} {\Delta\alpha_z}
= 
- \frac{\sin^2\theta}{(t_{\rm L} - t_{\rm R})\cos^2\theta}.
\end{eqnarray}

We can determine $t_1, t_2$, and $g$ by solving
the following three equations:
\begin{eqnarray}
a_L(t_1 - t_{\rm L})t_2 + a_R(t_2 - t_{\rm R})t_1 + 4g (a_L + a_R) &=& 0, \\
a_L(t_1-t_{\rm L}) + a_R(t_2 - t_{\rm R}) &=& 0, 
\end{eqnarray}
\begin{eqnarray}
2\sin\theta\left\{a_{\rm R}(t_2-t_{\rm R}) - a_{\rm L}(t_1-t_{\rm L})\right\}
 + \gamma^2(t_2-t_1)\left\{2(b_{\rm L} + b_{\rm R})\cos\theta +
		     (a_{\rm L} + a_{\rm R})\sin\theta\right\}
= 0.
\end{eqnarray}
When $t_{\rm L} - t_{\rm R} \ll 1$, 
$a_{\rm L}\simeq a_{\rm R}$ and $b_{\rm L}\simeq b_{\rm R}$.
Hence, we obtain
\begin{eqnarray}
t_1 &=& \frac{t_{\rm L} + t_{\rm R}}{2} + \frac{K}{2}(t_{\rm L} - t_{\rm R}),\\
t_2 &=& \frac{t_{\rm L} + t_{\rm R}}{2} - \frac{K}{2}(t_{\rm L} - t_{\rm R}),\\
g &=& -\frac{K(1-K)}{16}(t_{\rm L} - t_{\rm R})^2,
\end{eqnarray}
ignoring higher-order terms with respect to $t_{\rm L} - t_{\rm R}$.
Here,
\begin{eqnarray}
K &\coloneqq & 
\frac{c(t_{\rm L} - t_{\rm R})^2\cos^4\theta}
{c(t_{\rm L} - t_{\rm R})^2\cos^4\theta 
+ \sin^2\theta(2d\cos^2\theta + c\sin^2\theta)}
\end{eqnarray}
and
\begin{eqnarray}
c &\coloneqq& \frac{c_{\rm L} + c_{\rm R}}{2},\\
d &\coloneqq& \frac{d_{\rm L} + d_{\rm R}}{2},
\end{eqnarray}
where
\begin{eqnarray}
 c_\alpha &\coloneqq &
  \frac{\tilde{\Phi}_{\beta_\alpha}(2h)
  + \tilde{\Phi}_{\beta_\alpha}(-2h)}{2} \label{c's},\\
 d_\alpha &\coloneqq & \tilde{\Phi}_{\beta_\alpha}(0), \label{d's}
\end{eqnarray}
and $\alpha = {\rm L}, {\rm R}$. Note that $a_\alpha =
c_\alpha\sin\theta$ and $b_\alpha = d_\alpha\cos\theta$.
Since $c,d\ge 0$, $K$ satisfies $0\le K \le 1$. $K=0$ holds for $\theta=\pi/2$ 
(horizontal magnetic
field), and $K=1$ for $\theta = 0$ (vertical magnetic field).

\end{document}